\documentclass[a4paper]{article}
\usepackage{INTERSPEECH2020}
\usepackage{times}
\usepackage{epsfig}
\usepackage{graphicx}
\usepackage{epstopdf}
\usepackage{amsmath}
\usepackage{amssymb}
\usepackage{mathtools}
\usepackage{multirow}
\usepackage{enumerate}
\usepackage{enumitem}
\usepackage{makecell}
\usepackage{url}
\usepackage[noend]{algpseudocode}
\usepackage[ruled,vlined]{algorithm2e}
\usepackage{verbatim}
\usepackage[keeplastbox]{flushend}
\usepackage[utf8]{inputenc}
\usepackage[english]{babel}
\usepackage{placeins}
\usepackage{color}
\usepackage{soul}
\usepackage{hyperref}
\usepackage{subcaption}
\usepackage{icomma} 

\hypersetup{
    colorlinks=true,
    linkcolor=blue,
    filecolor=magenta,      
    urlcolor=cyan,
}

\graphicspath{ {Figures/} }

\title{JukeBox: A Multilingual Singer Recognition Dataset}
\name{Anurag Chowdhury, Austin Cozzo, Arun Ross}
\address{Michigan State University}
\email{\{chowdh51, cozzoaus, rossarun\}@cse.msu.edu}

\begin{document}

\maketitle
\begin{abstract}
A text-independent speaker recognition system relies on successfully encoding speech factors such as vocal pitch, intensity, and timbre to achieve good performance. A majority of such systems are trained and evaluated using spoken voice or everyday conversational voice data. Spoken voice, however, exhibits a limited range of possible speaker dynamics, thus constraining the utility of the derived speaker recognition models. Singing voice, on the other hand, covers a broader range of vocal and ambient factors and can, therefore, be used to evaluate the robustness of a speaker recognition system. However, a majority of existing speaker recognition datasets only focus on the spoken voice. In comparison, there is a significant shortage of labeled singing voice data suitable for speaker recognition research. To address this issue, we assemble \textit{JukeBox} - a speaker recognition dataset with multilingual singing voice audio annotated with singer identity, gender, and language labels. We use the current state-of-the-art methods to demonstrate the difficulty of performing speaker recognition on singing voice using models trained on spoken voice alone. We also evaluate the effect of gender and language on speaker recognition performance, both in spoken and singing voice data. The complete \textit{JukeBox} dataset can be accessed at http://iprobe.cse.msu.edu/datasets/jukebox.html.

\end{abstract}

\vspace{-0.2cm}
\section{Introduction}
Speaker recognition entails comparing two audio samples encompassing human voice and determining if the voices pertain to the same individual. A majority of speaker recognition research has focused on modeling the speaker-dependent characteristics from conversational or spoken voice data~\cite{kinnunen2010overview}. However, the spoken voice only exhibits a limited range of possible speaker dynamics~\cite{sundberg1977acoustics}. As a result, such speaker recognition systems generalize poorly to a wide variety of speaking styles and vocal effort~\cite{shriberg2008effects}. The singing voice is one such example of a speaking style~\cite{Mehrabani2013SingerClusteringGMM}, where the speaker-dependent voice characteristics depart heavily from the spoken voice of the same speaker. Apart from the perceived differences in intensity, pitch, and timbre, there are also differences in the physiological formation of sung speech~\cite{daniloff1994allophonic}, especially when considering a trained singer~\cite{Brown1988PhysiologicalDB}. The different styles of singing further diversify the acoustic differences between spoken and singing speech~\cite{Stone2003Acoustic}, leading to several challenges for speaker recognition systems. One of the primary challenges of speaker recognition from singing is the increased intra-user variance and decreased inter-user variance due to intentional voice modulation, across a broad acoustic spectrum~\cite{sundberg1977acoustics}. In addition, the presence of background music and chorus increases the challenges of the task. Thus, a speaker recognition system's ability to correctly match a singer's voice across multiple songs can be used to assess its robustness.

However, there appears to be limited amount of work done on this topic. Some of the relevant early literature treat singing voice as a speaking style and cluster it using speaker clustering algorithms~\cite{Mehrabani2013SingerClusteringGMM,Mehrabani2012SingerClustering}. In another work~\cite{Patil2012HumSingSpeak}, the authors use singing voice to perform speaker recognition; however, no cross-modal experiments were done, i.e. training a model on speaking data and testing on singing data (or vice versa). This work was extended in~\cite{Chhayani2013CrossModality} to evaluate cross-modal speaker recognition; however, poor performance was reported. Notably, the datasets used in~\cite{Mehrabani2013SingerClusteringGMM,Mehrabani2012SingerClustering, Patil2012HumSingSpeak, Chhayani2013CrossModality} were limited to a small set ($\leq 50$) of speakers.

One key reason behind the underrepresented research focus on speaker recognition from singing voice, i.e., singer recognition, is the lack of sufficient development and evaluation data. A review of currently existing music datasets for research (in Table~\ref{tab:spkr_related}) reveals two relevant datasets: the Million Song Dataset (MSD)~\cite{Bertin2011MSD} and the Free Music Archive (FMA)~\cite{Defferrard2017FMA}. MSD contains $1,000,000$ songs from $44,745$ artists/groups. However, the data is available only in the form of audio features and not raw audio, which forces a speaker recognition algorithm to work with a predetermined feature-set. FMA, on the other hand, contains $106,574$ songs from $16,341$ artists/groups. Here, the `artist/group' label refers to the associated music group/band and not necessarily the individual singer, who might change over time. For example, both Ozzy Osbourne and Ronnie James Dio have sung songs under the artist label of Black Sabbath, thus making group/band labels unsuitable for training or testing a speaker recognition system.

Therefore, in this work, we assemble \textit{JukeBox}, a singing voice dataset annotated with singer, gender, and language labels for the development and evaluation of speaker recognition methods. In the next few sections, we will describe in detail this dataset, the data collection procedure, several experimental protocols, and analyze the performance of state-of-the-art speaker recognition methods on the dataset.

\begin{table}[t]
  \fontsize{7}{9}\selectfont
  \caption{A list of related music datasets compared to the \textit{JukeBox} dataset.}
  \centering
\resizebox{\columnwidth}{!}{%
\begin{tabular}{|c|c|c|c|c|}
\hline
Dataset & \makecell{Number\\ of Samples} & \makecell{Number\\ of Artists} & Label  & \makecell{Raw\\ Audio}\\
\hline\hline
UT-Sing~\cite{Mehrabani2012SingerClustering} & 165 & 33 & Singer & Yes \\
\hline
MusiClef~\cite{Schedl2013MusiClef} & 1,355 & 218 & Artist / Group & No\\
\hline
Homburg~\cite{Homburg2005} & 1,886 & 1,463 & Artist / Group & Yes\\
\hline
1517-Artists~\cite{Seyerlehner20081517} & 3,180 & 1,517 & Artist / Group & Yes\\
\hline
Unique~\cite{Seyerlehner2010Unique} & 3,115 & 3,115 & Artist / Group & Yes\\
\hline
USPOP~\cite{Berenzweig2003USPOP} & 8,752 & 400 & Artist / Group & No\\
\hline
CAL10K~\cite{Tingle2010CAL10k} & 10,271 & 4,597 & Artist / Group  & No\\
\hline
MagnaTagATune~\cite{Law2009TagATune} & 16,389 & 270 & Artist / Group & Yes\\
\hline
Codiach~\cite{Mckay06alarge} & 20,849 & 1,941 & Artist / Group & No\\
\hline
FMA~\cite{fma_dataset} & 106,574 & 16,341 & Artist / Group & Yes\\
\hline
OMRAS2~\cite{mauch2009} & 152,410 & 6,983 & Artist / Group & No\\
\hline
MSD~\cite{Bertin2011MSD} & 1,000,000 & 44,745 & Artist / Group & No\\
\hline
\textbf{\textit{JukeBox}} & \textbf{7,000} & \textbf{936} & \textbf{Singer} & \textbf{Yes}\\
\hline
\end{tabular}%
}

\label{tab:spkr_related}
\end{table}

\section{\textit{JukeBox} Dataset}~\label{sec:dataset}
\begin{figure}[t]		
		\centering
		\includegraphics[width=.45\textwidth]{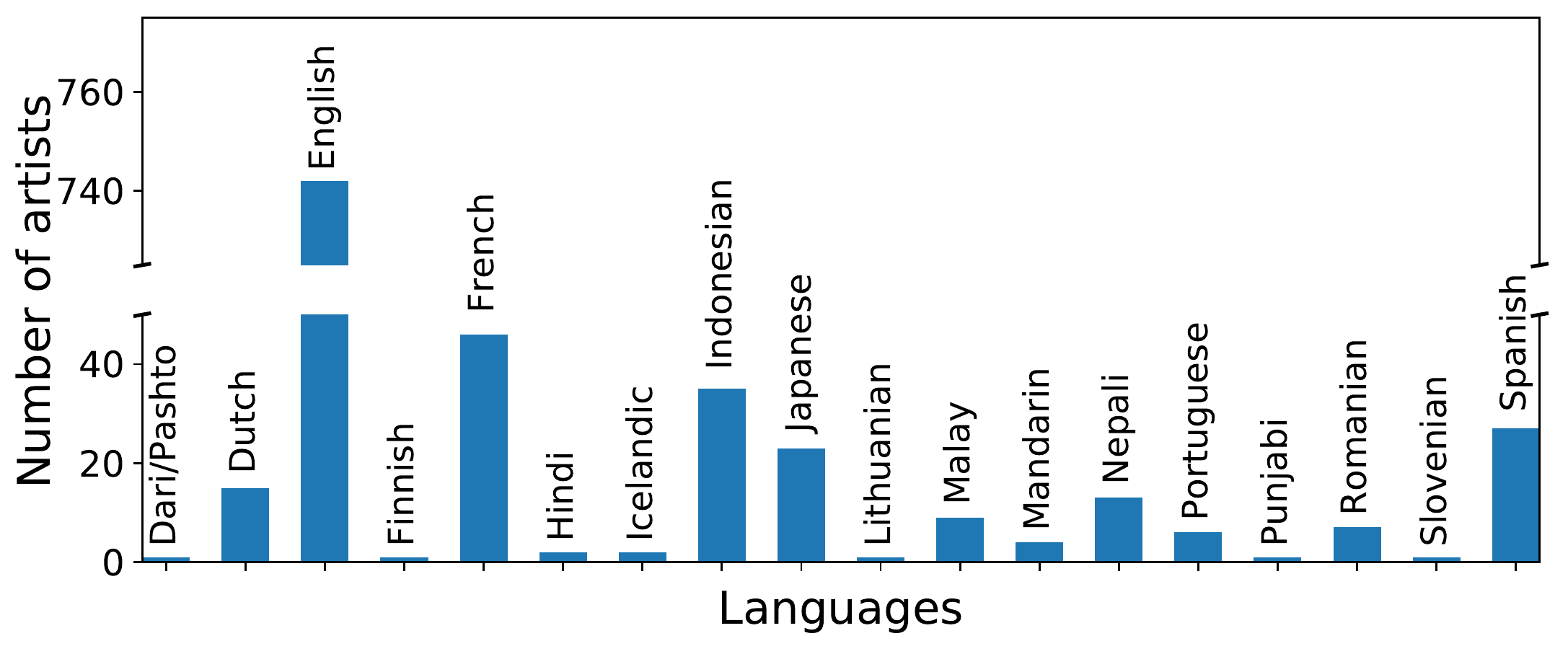}
		\caption{Distribution of languages in the \textit{JukeBox} dataset}
		\label{fig:lang_bar}
	\end{figure}

\begin{figure}[t]		
		\centering
		\includegraphics[width=.45\textwidth]{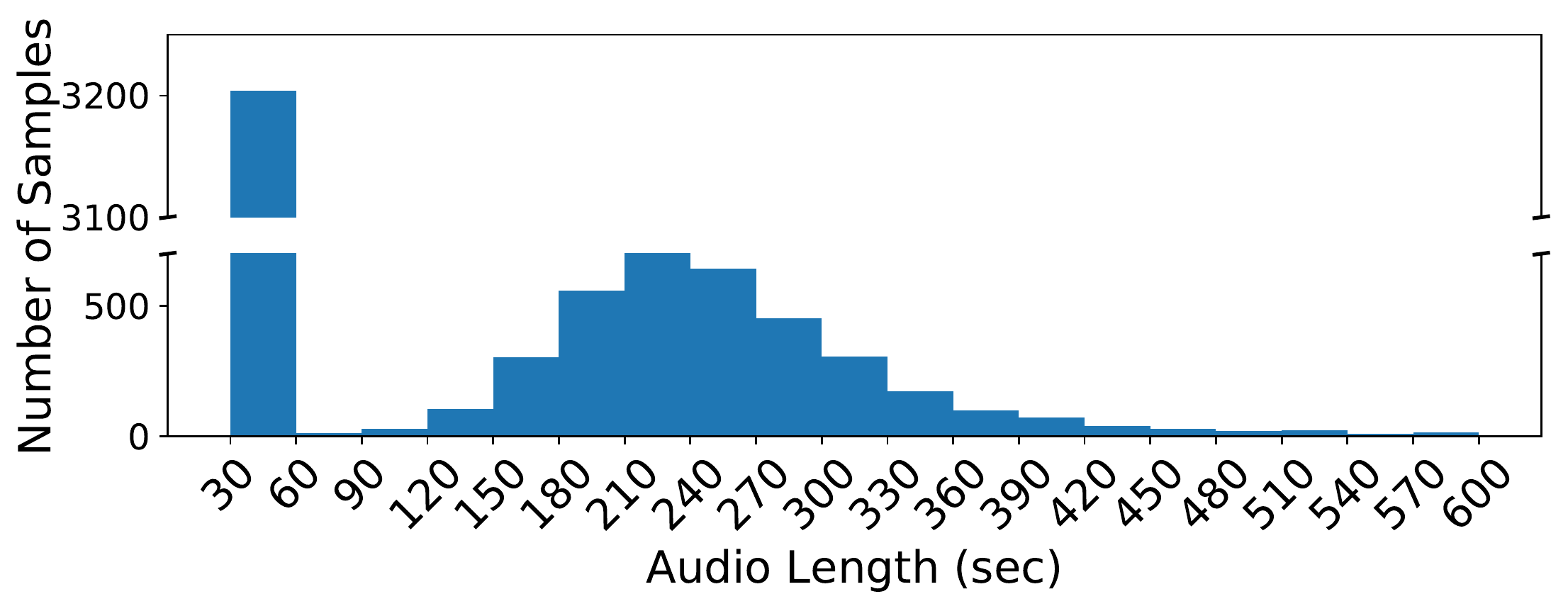}
		\caption{Distribution of audio length in the \textit{JukeBox} dataset}
		\label{fig:audio_hist}
		\vspace{-0.5cm}
	\end{figure}

The \textit{JukeBox} dataset contains $467$ hours of singing audio data sampled at $16$ KHz, downloaded from the Internet Archive (IA)~\cite{internet_archive}. There is a total of $936$ different singers in the dataset, of which $533$ are male. Figures~\ref{fig:lang_bar} and ~\ref{fig:audio_hist} summarize the different languages and the distribution of the length of songs in the \textit{JukeBox} dataset. The songs in the \textit{JukeBox} dataset:

\begin{itemize}[leftmargin=0cm,itemindent=.5cm,labelwidth=\itemindent,labelsep=0cm,align=left]
\setlength\itemsep{0em}
\item are sung in $18$ different languages, as shown in Figure~\ref{fig:lang_bar}, where almost one-fifth of the singers in the dataset sing in non-English languages (i.e., a language other than English).
\item are recorded under a wide variety of acoustic environments and recording apparatus, ranging from highly-constrained studio recording setups to completely-unconstrained live concerts.
\item contain multiple singers apart from the person-of-interest (POI), for example, vocal duets with overlapped singing and background chorus.
\item  contain different types of background music (such as drums, piano, or other instrumentation), thus adding to the difficulty of performing speaker recognition.
\end{itemize}

\subsection{Data collection procedure}
The \textit{JukeBox} dataset was assembled as follows.
\begin{itemize}[leftmargin=0cm,itemindent=.5cm,labelwidth=\itemindent,labelsep=0cm,align=left]
\setlength\itemsep{0em}
\item \textbf{Candidate list creation for artists of interest:}
We started by compiling a list of artists from Wikipedia, who were tagged as ``singer". This yielded   a list of $5,046$ artists of interest (AOI) from a variety of languages and genres (such as Pop, R\&B, Rock, Jazz, Folk, Classical, etc.), with associated metadata such as country of origin ($\sim18$ different countries) and years active.

\item \textbf{Candidate list creation for songs of interest:}
The candidate list for AOI was used to query Spotify's song database~\cite{spotify_api} to generate a list of $162,311$ songs. This list was then cross-referenced against IA's repository to generate a list of downloadable songs of interest (SOI). We chose IA as our audio source due to its (a) large collection of audio, (b) public accessibility, (c) nearly unrestricted download access~\cite{internet_archive_api}, and (d) re-distribution permission for non-commercial purposes. 

\item \textbf{Downloading songs of interest:}
The IA repository often contains multiple copies of a song, differing in their audio duration, recording conditions (such as studio versus live versions), and singers (such as original versus cover artists). We specifically avoided cover artists to remove multiple versions of a song and ensure the correctness of artist labels. A large number of the songs on IA were restricted to $30$-second duration due to copyright concerns. We preferred the full duration versions of a song, whenever available. Using these criteria, we downloaded a total of $10,063$ SOI for $1,341$ AOI.

\item \textbf{SOI pruning for removing non-singing audios:} Voice Activation Detection (VAD)~\cite{webrtc_vad} was used on the SOI to remove silent segments. The VAD processed songs were then manually verified to discard audio files that did not contain singing vocals. Note that the human listeners only listened to 5 equally separated 1-second long audio segments in every song to make their decision. This process ensured a practicable manual verification process of $~1,500$ hours of audio data. 


\item \textbf{Manual verification of language labels in non-English songs:} Nearly one-fifth of the singers in the \textit{JukeBox} dataset are non-English singers. The language labels originally assumed the non-English singers to sing in a non-English language. However, some of the non-English singers were multilingual, and had songs in the English language as well. Therefore, a secondary manual verification of the dataset was conducted to remove English songs for non-English singers. The resulting $7,000$ SOI from $936$ AOI form the \textit{JukeBox} dataset.

\item \textbf{Splitting the dataset into the train, test, and auxiliary subsets:}
Finally, the set of $936$ speakers in the dataset was split into three subsets (shown in Table~\ref{tab:dataset_stats}):
\vspace{-0.1cm}
\begin{itemize}[leftmargin=0cm,itemindent=.5cm,labelwidth=\itemindent,labelsep=0cm,align=left]
\setlength\itemsep{0em}
\item Training set: All speakers with at least three audio samples constitute the training set ($670$ subjects). This set is reserved for training or fine-tuning speaker recognition models.

\item Test set: All speakers with exactly two audio samples constitute the test set ($98$ subjects). This set is reserved for evaluating trained speaker recognition models on singing voice data.

\item Auxiliary set: All speakers with only one audio sample constitute the auxiliary set ($168$ subjects).
This set can be used to augment the training data for speaker recognition models trained in the identification mode. However, the auxiliary set cannot be used to train models in the verification mode, as at least 2 samples per subject are needed to form a genuine pair.

\end{itemize}
\vspace{-0.5cm}

\end{itemize}

\begin{table}[ht]
	\fontsize{8}{10}\selectfont
    \centering
    \caption{Dataset statistics of the \textit{JukeBox} dataset}
    \label{tab:dataset_stats}
    \begin{tabular}{lrrr}
        \textbf{Dataset} & \textbf{Train} & \textbf{Test} & \textbf{Auxillary}\\
        \hline
        \# of Subjects & 670 & 98 & 168\\
        \# of Male Subjects & 397 & 57 & 79\\
        \# of Non-English Subjects & 104 & 21 & 69  \\
        \# of Samples & 6,636 & 196 &  168\\
        \# of Hours & 385 & 33 & 49\\
        Max \# of Samples/Speaker & 87 & 2 & 1\\
        Min \# of Samples/Speaker & 3 & 2 & 1\\
        Avg \# of Samples/Speaker & 10 & 2 & 1
    \end{tabular}
    \vspace{-0.5cm}
\end{table}

\section{Datasets and Experimental Protocols}
We propose several experimental protocols for establishing baseline speaker recognition performance on the \textit{JukeBox} dataset. We use state-of-the-art and baseline speaker recognition methods, viz., 1D-Triplet-CNN~\cite{chowdhury2020fusing}, xVector-PLDA~\cite{snyder2018x}, and iVector-PLDA~\cite{garcia2011analysis} for this purpose. We also evaluate their performance on the \textit{JukeBox} dataset under different conditions based on gender of the artists and language of the songs.

\subsection{Datasets}
\vspace{-0.2cm}
\subsubsection{VoxCeleb2 Dataset}
We use the VoxCeleb2~\cite{chung2018voxceleb2} dataset to perform baseline speaker recognition experiments on spoken voice data (i.e. spoken-to-spoken scenario). We use a subset of the VoxCeleb2 dataset to keep the experiments computationally tractable. A random subset of $5,994$ video samples corresponding to the $5,994$ celebrities in the VoxCeleb2 dataset forms the training set. Similarly, a random subset of $118$ video samples corresponding to $118$ celebrities forms the evaluation set. Speech from each video in the dataset is extracted and split into multiple non-overlapping 5-second long audio samples.
\vspace{-0.3cm}
\subsubsection{\textit{JukeBox} Dataset}
Data from \textit{JukeBox} dataset is used to \textit{fine-tune} and evaluate the aforementioned speaker recognition methods on singing voice data (i.e. both spoken-to-singing and singing-to-singing scenarios). Each song in the training set was split into multiple non-overlapping $30$-second long segments to increase the number of training samples. In all our experiments, we use the samples from the training set to train the speaker verification algorithms, and the samples from the test set to evaluate the performance of the trained speaker verification models.

\subsection{Experimental Protocol}~\label{sec:experiments}
\vspace{-0.8cm}
\subsubsection{iVector-PLDA based speaker verification experiments}
We use the MSR Identity Toolkit's~\cite{sadjadi2013msr} implementation of the iVector-PLDA algorithm as our first baseline speaker verification method.
A Gaussian-PLDA (gPLDA)-based matcher~\cite{sadjadi2013msr} is used to compare the extracted i-Vector embeddings of a pair of speech samples.
\vspace{-0.3cm}    
\subsubsection{xVector-PLDA based speaker verification experiments}
We use the PyTorch-based implementation~\cite{chowdhury2020fusing} of the xVector algorithm as our second baseline speaker verification method. A gPLDA-based matcher~\cite{sadjadi2013msr} is used to compare the extracted xVector embeddings of a pair of speech samples.

\begin{figure*}[ht]
\begin{subfigure}{.33\textwidth}
  \centering
  \includegraphics[width=1\linewidth]{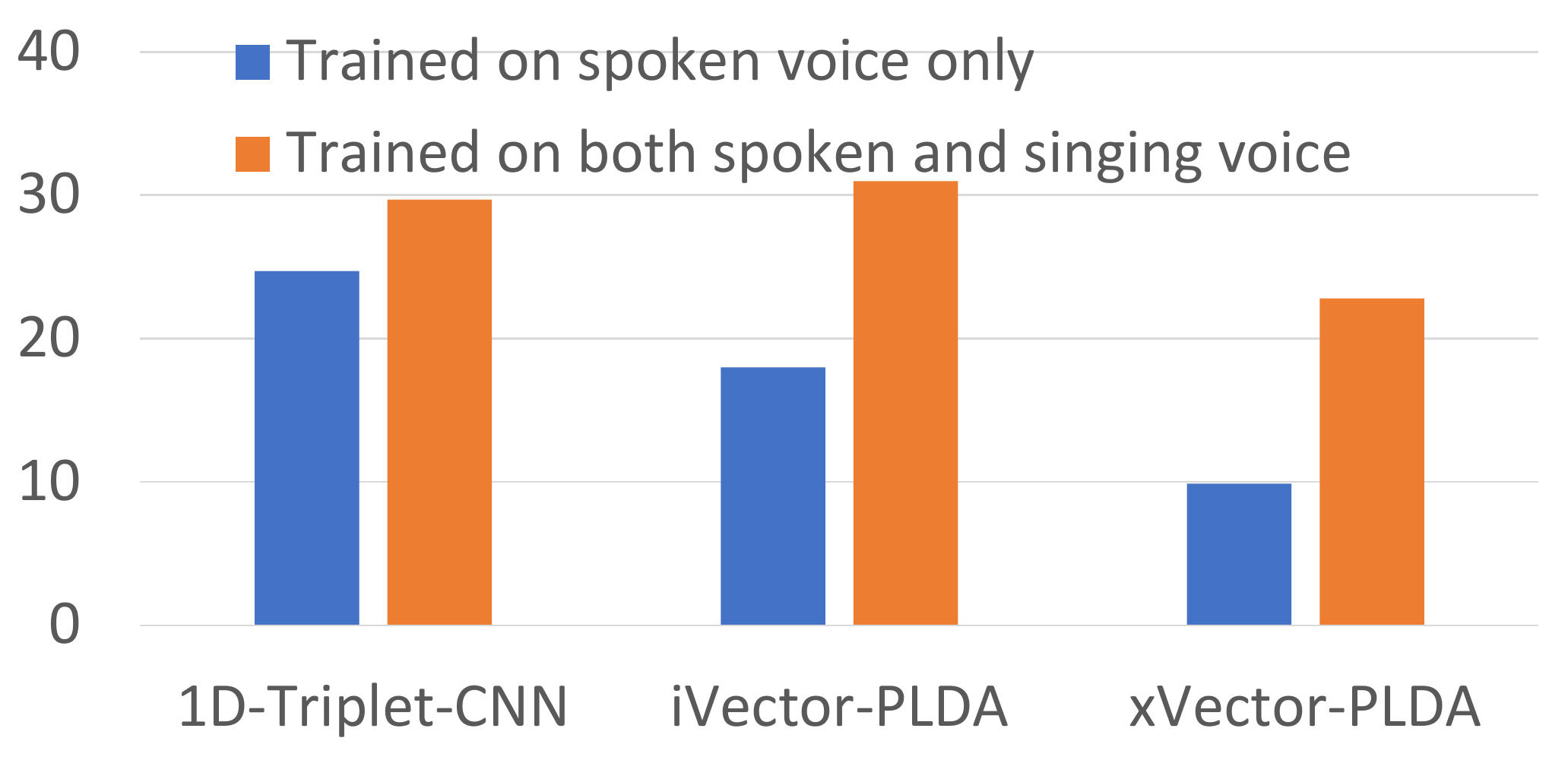}  
  \caption{Effect of training data on singer verification.}
  \label{fig:sub-first}
\end{subfigure}
\begin{subfigure}{.33\textwidth}
  \centering
  \includegraphics[width=1\linewidth]{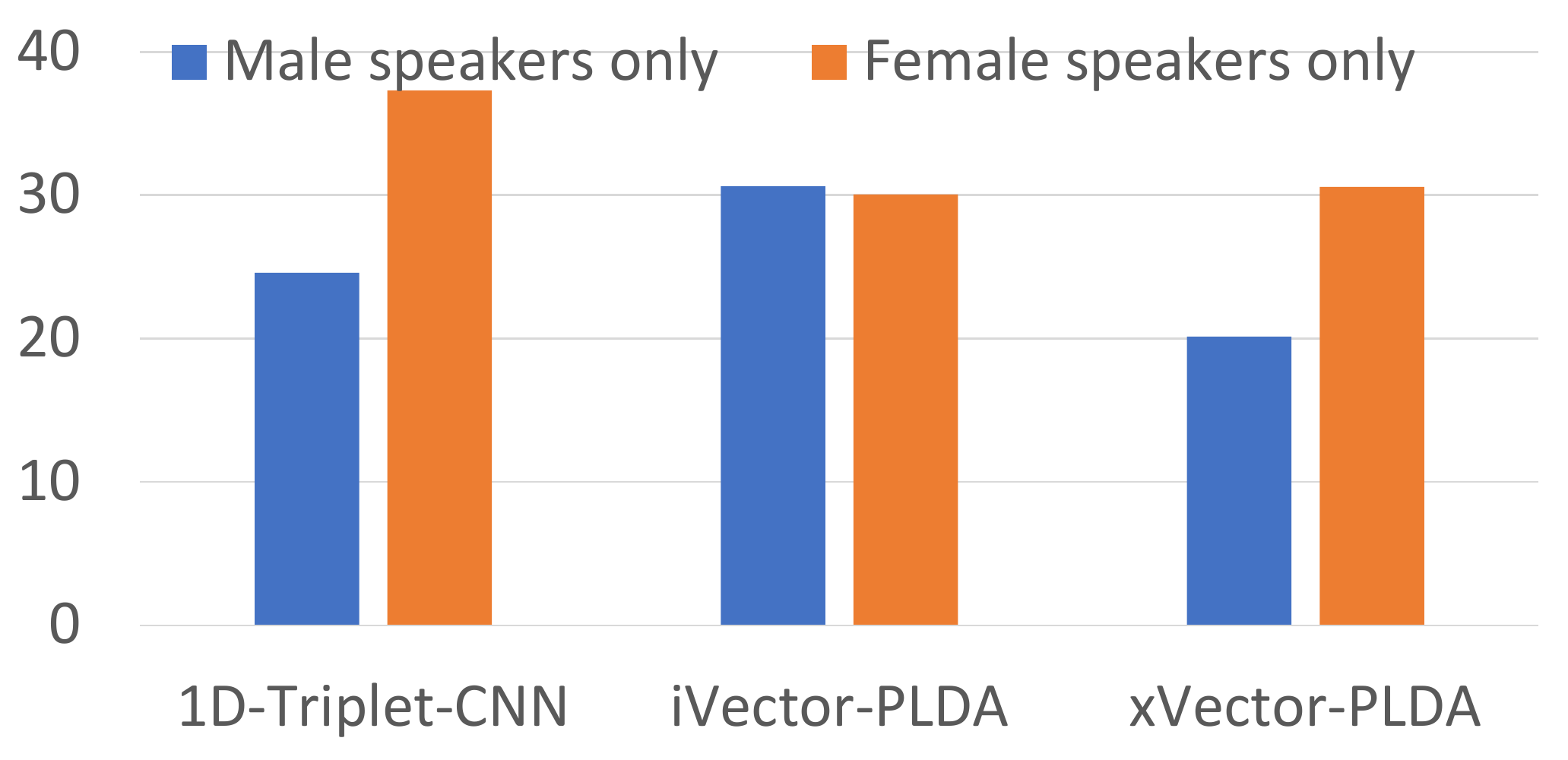}  
  \caption{Effect of gender on singer verification.}
  \label{fig:sub-second}
\end{subfigure}
\begin{subfigure}{.33\textwidth}
  \centering
  \includegraphics[width=1\linewidth]{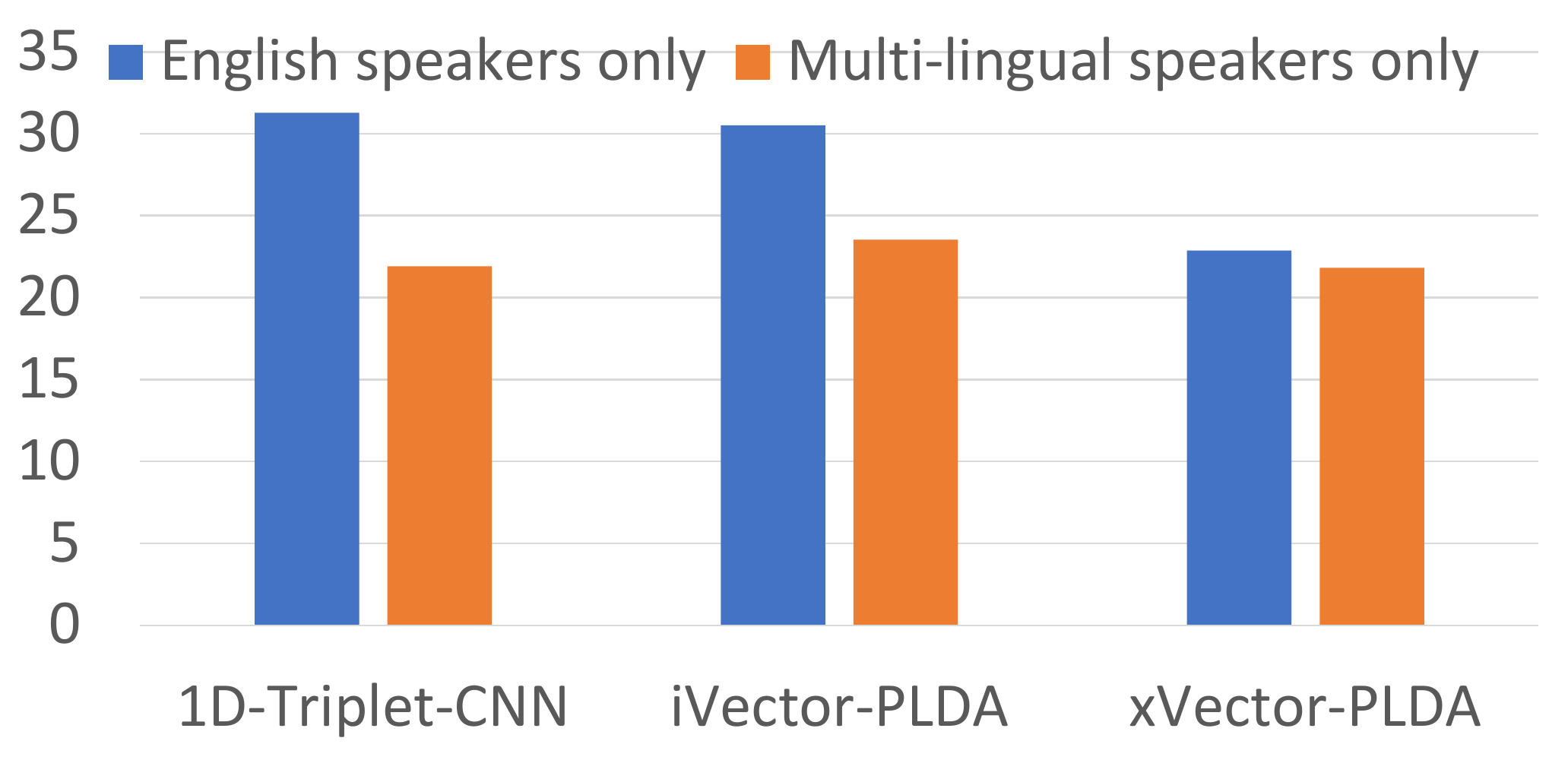}  
  \caption{Effect of language on singer verification.}
  \label{fig:sub-third}
\end{subfigure}
\caption{Summary of verification performance (TMR@FMR=$1\%$) across different evaluation conditions on the \textit{JukeBox} dataset.}
\label{fig:analysis}
\vspace{-0.3cm}
\end{figure*}

\vspace{-0.3cm}
\subsubsection{1D-Triplet-CNN based speaker verification experiments}
We also perform speaker verification experiments using the 1D-Triplet-CNN algorithm, due to its demonstrated robustness to audio degradations~\cite{chowdhury2020fusing}. The audio samples in the training set are grouped into triplets to train the 1D-Triplet-CNN algorithm. For evaluation, the audio samples are grouped into pairs and processed by the trained model to generate pairs of 1D-Triplet-CNN embeddings. These pairs of embeddings are then matched using the cosine similarity metric.

\begin{table}[t]
	\fontsize{7}{9}\selectfont
	\caption{Speaker verification results on spoken voice data from the VoxCeleb2 dataset using the 1D-Triplet-CNN~\textbf{[M1]}, iVector-PLDA~\textbf{[M2]}, and xVector-PLDA~\textbf{[M3]} models. The same models are evaluated on the \textit{JukeBox} dataset to compare the performance on singing voice data. Here, \textbf{P1} = VoxCeleb2 , \textbf{P2} = \textit{JukeBox} , and \textbf{P3} = Both VoxCeleb2 and \textit{JukeBox} together.}
	\centering
    \begin{tabular}{|c|c|c|c|c|c|}
    \hline 
    Exp. \# & \makecell{Train Set\\ /Test Set} & Models &\makecell{TMR\\@FMR=1\%} & minDCF & \makecell{EER\\(in \%)}\tabularnewline
    
    \hline 
    \hline
    1 & \multirow{4}{*}{P1/P1} &M1 & 91.23 & 1.82 & 4.09\tabularnewline
   \cline{1-1} \cline{3-6} 
    2 & & M2 & 92.79 & 1.38 & 3.81\tabularnewline
    \cline{1-1} \cline{3-6}
    3 & & M3 & 65.06 & 4.15 & 7.89\tabularnewline
    \hline 
    4 & \multirow{4}{*}{P1/P2} & M1 & \textbf{24.72} & \textbf{8.35} & 26.48\tabularnewline
    \cline{1-1} \cline{3-6}
    5 & & M2 & 18 & 8.99 & \textbf{24.49}\tabularnewline
    \cline{1-1} \cline{3-6} 
    6 & & M3 & 9.9 & 9.56 & 31.83\tabularnewline
    \cline{1-2} \cline{3-6} 
    7 & \multirow{4}{*}{P3/P2} &M1 & 29.71 & 7.91 & 24.36\tabularnewline
    \cline{1-1} \cline{3-6}  
    8 & & M2 & \textbf{30.98} & \textbf{7.77} & \textbf{23.63}\tabularnewline
    \cline{1-1} \cline{3-6} 
    9 & & M3 & 22.82 & 8.42 & 26.39\tabularnewline
    \hline 
   
    \end{tabular}

    \vspace{-0.2cm}
	
	\label{tab:spkr_veri}
\end{table}
	
\begin{table}[t]
	\fontsize{7}{9}\selectfont
	\caption{Verification results on the gender and language specific evaluation subsets of the \textit{JukeBox} dataset using the 1D-Triplet-CNN~\textbf{[M1]}, iVector-PLDA~\textbf{[M2]}, and xVector-PLDA~\textbf{[M3]} methods. All the models were trained on the VoxCeleb2 dataset and fine-tuned using the \textit{JukeBox} dataset. Here, \textbf{C1} = male speakers only, \textbf{C2} = female speakers only, \textbf{C3} = English speakers only, and \textbf{C4} = non-English speakers only.}
	\centering	
    \begin{tabular}{|c|c|c|c|c|c|}
    \hline 
    Exp. \# & Models & \makecell{Evaluation\\Condition} & \makecell{TMR\\@FMR=1\%} & minDCF & \makecell{EER\\(in \%)}\tabularnewline
    \hline 
    \hline 
    10 & \multirow{4}{*}{M1} & C1 & 24.6 & 8.33 & 24.44\tabularnewline
    \cline{1-1} \cline{3-6}
    11 & & C2 & 37.29 & 6.4 & 21.95\tabularnewline
     \cline{1-1} \cline{3-6} 
    12 & & C3 & 31.28 & 7.67 & 21.7\tabularnewline
     \cline{1-1} \cline{3-6}
    13 & & C4 & 21.91 & 8.18 & 33.63\tabularnewline
    \hline 
    14 & \multirow{4}{*}{M2} & C1 & 30.64 & 7.87 & 26.41\tabularnewline
    \cline{1-1} \cline{3-6} 
    15 & & C2 & 30.05 & 7.58 & 22.43\tabularnewline
    \cline{1-1} \cline{3-6}  
    16 & & C3 & 30.51 & 7.75 & 23.67\tabularnewline
    \cline{1-1} \cline{3-6} 
    17 & & C4 & 23.53 & 7.67 & 28.48\tabularnewline
    \hline 
    18 & \multirow{4}{*}{M3} & C1 & 20.14 & 8.57 & 25.09\tabularnewline
    \cline{1-1} \cline{3-6}  
    19 & & C2 & 30.59 & 7.72 & 29.29\tabularnewline
    \cline{1-1} \cline{3-6}  
    20 & & C3 & 22.88 & 8.41 & 24.72\tabularnewline
    \cline{1-1} \cline{3-6}  
    21 & & C4 & 21.81 & 8.44 & 38.96\tabularnewline
    \hline 
    \end{tabular}
    
    \vspace{-0.3cm}
	
	\label{tab:spkr_ver_gen_lang}
\end{table}

\vspace{-0.3cm}
\subsubsection{Studying the effect of gender on speaker verification}~\label{sec:effect_of_gender}
The fundamental physiological differences between male and female voices~\cite{mason1993gender} have been used to advocate for their separate treatment in the context of speaker recognition~\cite{li2015gender}. These differences are further pronounced in the singing voice~\cite{Sulter1996MaleFemalePhonetograms}. Male singers, for example, exhibit a larger variation in their falsetto (a method of voice production)~\cite{Welch1988MaleFalsetto}, potentially making them harder to recognize than their female counterparts. Therefore, in this work, we perform gender-specific speaker verification experiments (Exp. \# 10, 11, 14, 15, 18, and 19 in Table~\ref{tab:spkr_ver_gen_lang}) to study the effect of gender on speaker verification from singing voice data. We use the following two types of gender-specific trials in our experiments:

\vspace{-0.2cm}
\begin{itemize}[leftmargin=0cm,itemindent=.5cm,labelwidth=\itemindent,labelsep=0cm,align=left]
\setlength\itemsep{0em}
    \item[\textbf{Female only trials:} ] In these experiments, the trained models are evaluated on same-gender (female only) trials drawn from $41$ female artists in the test set of the \textit{JukeBox} dataset. 
    \item[\textbf{Male only trials:} ] In these experiments, the trained models are evaluated on same-gender (male only) trials drawn from $57$ male artists in the test set of the \textit{JukeBox} dataset. 
\end{itemize}

\vspace{-0.3cm}
\subsubsection{Studying the effect of language on speaker verification}

Speaker recognition performance of both humans and machines degrade when the speech audio being evaluated is in a language unknown or unfamiliar to the listener~\cite{lu2009effect}. This is also known as the language-familiarity effect (LFE)~\cite{fleming2014language}. In this work, we perform additional speaker verification experiments on the \textit{JukeBox} dataset to evaluate the effect of language on speaker verification performance from singing audio. We perform two different types of language-based speaker verification experiments, given by Exp. \# 12, 13, 16, 17, 20, and 21 in Table~\ref{tab:spkr_ver_gen_lang} and described below. All the models in this set of experiments were trained and fine-tuned using the multilingual speech data from the VoxCeleb2 and the \textit{JukeBox} datasets, respectively. 
\vspace{-0.1cm}
\begin{itemize}[leftmargin=0cm,itemindent=.5cm,labelwidth=\itemindent,labelsep=0cm,align=left]
    \setlength\itemsep{0em}
    \item[\textbf{Same language, English only trials:} ] In these experiments, the models are evaluated on same-language (English only) trials drawn from $77$ English singers in the test set of \textit{JukeBox}. 
    
    \item[\textbf{Multilingual, non-English trials:} ] In these experiments, the models are evaluated on multilingual trials drawn from $21$ non-English singers in the test set of \textit{JukeBox}. The songs in the multilingual trials are sung in one of these $9$ different non-English languages: Dari/Pashto, Dutch, French, Japanese, Mandarin, Nepali, Punjabi, Romanian, Spanish.
\end{itemize}

\vspace{-0.3cm}
\subsubsection{Studying the effect of singing style modeling on speaker verification}

Finally, we also perform a fusion of Global Style Token (GST)~\cite{wang2018style} based prosodic speech features with the 1D-Triplet-CNN based speaker embedding to facilitate singing style modeling for speaker verification. In these experiments, we extract the speaker embeddings obtained from the 1D-Triplet-CNN and input it to GST to extract prosodic speech features. These prosodic speech features are further fused with the 1D-Triplet-CNN based speaker embeddings to derive a style-sensitive speaker embedding. This embedding is then used to perform speaker verification experiments, given in Table~\ref{tab:spkr_veri_prosody}.

\begin{table}[t]
	\fontsize{7}{9}\selectfont
	\caption{Effect of prosody modeling for singing-style based speaker recognition. The 1D-Triplet-CNN + GST model performs singing-style based speaker recognition. The numbers represent performance when trained on the VoxCeleb2 dataset only / on both the VoxCeleb2 and the \textit{JukeBox} datasets}
	\centering
    \begin{tabular}{|c|c|c|c|}
    \hline 
    Models & TMR@FMR=1\% & minDCF & EER (in \%)\tabularnewline
    \hline 
    \hline 
    1D-Triplet-CNN & \textbf{24.72/29.71} & \textbf{8.35/7.91} & \textbf{26.48}/24.36\tabularnewline
    \hline
    1D-Triplet-CNN + GST & 19.42/26.80 & 8.78/8.24 & 26.55/\textbf{24.27}\tabularnewline
    \hline 
    \end{tabular}
	\vspace{-0.3cm}
	\label{tab:spkr_veri_prosody}
\end{table}

\section{Results and Analysis}
	The results of all the experiments described in Section~\ref{sec:experiments} are given in Tables~\ref{tab:spkr_veri}, \ref{tab:spkr_ver_gen_lang}, and \ref{tab:spkr_veri_prosody}, and Figure~\ref{fig:analysis}. For all the speaker verification experiments, we report the
	True Match Rate at a False Match Rate of $1\%$ (TMR@FMR=$1\%$), minimum Detection Cost Function (minDCF) and Equal Error Rate (EER in $\%$). The minimum Detection Cost Function (minDCF) is computed at a prior probability of $0.01$ for the specified target speaker ($P_{tar}$) with a cost of missed detection of $10$ ($C_{miss}$).
	
	\begin{itemize}[leftmargin=0cm,itemindent=.5cm,labelwidth=\itemindent,labelsep=0cm,align=left]
	\setlength\itemsep{0em}
	
	\item In the experiments 1 to 3 given in Table~\ref{tab:spkr_veri},  baseline speaker verification performance is established for all the models on spoken voice data from the VoxCeleb2 dataset. The relatively lower performance of the xVector-PLDA model is attributed to the limited training data being insufficient for learning xVector-PLDA model's considerably larger parameter space.
	
	\item Further, in experiments 1 to 6, a large performance drop is noted across all models when they are evaluated on the \textit{JukeBox} dataset when compared to the VoxCeleb2 dataset. This indicates the difficulty of performing singer recognition using models that are pre-trained on spoken voices. 
	
	\item Fine-tuning the models pre-trained on the VoxCeleb2 dataset, using the training set of \textit{JukeBox} (in experiments 7 to 9) improved the average performance (TMR@FMR=$1\%$) of all the models by $\sim 10.29\%$. This indicates the benefit of using \textit{JukeBox} for fine-tuning pre-trained speaker recognition models for the task of singer recognition.
	
	\item We also performed speaker identification experiments corresponding to the experimental protocol given in Table~\ref{tab:spkr_veri}. The identification results follow the trend seen in verification. Best performance is observed when the models are trained and tested on spoken voice. Worst performance is observed when the models are trained on spoken voice and tested on singing voice. Fine-tuning the models trained on spoken voice with singing voice improves the performance on singing voice.

	\item In the gender-based speaker verification experiments (10, 11, 14, 15, 18, and 19) given in Table~\ref{tab:spkr_ver_gen_lang}, majority of the models perform better on female subjects. This is an interesting result because (a) both the VoxCeleb2 and \textit{JukeBox} datasets have a higher proportion of male subjects in the training data, and (b) gender-based speaker recognition experiments on spoken speech data usually perform better for males~\cite{mason1993gender,li2015gender}. This demonstrates the effect of gender-specific voice range profiles of the singing voice~\cite{Sulter1996MaleFemalePhonetograms} in the context of speaker recognition.
	
	\item In the language-based speaker verification experiments (12, 13, 16, 17, 20, and 21) given in Table~\ref{tab:spkr_ver_gen_lang}, majority of the models perform better on English-only trials. This indicates the presence of the LFE even in singing audios, where the speaker models trained on English-majority speech data performs better on English-only speech data compared to non-English speech.
	
	\item The inclusion of prosody modeling for encoding the singing style in the speaker embeddings degrades the speaker verification performance (see Table~\ref{tab:spkr_veri_prosody}). This can be attributed to the large intra-speaker variance due to different singing styles used in different songs. This indicates that the singing-style of the singer estimated from a fixed set of songs does not generalize well across other songs, leading to a drop in performance.

	\end{itemize}
	
\vspace{-0.2cm}
\section{Summary}
We assembled a multilingual singer recognition dataset called \textit{JukeBox}. The evaluation of state-of-the-art speaker recognition methods trained only on spoken voice data, on the \textit{JukeBox} dataset, revealed the challenges posed by singing voice data to speaker recognition. The \textit{JukeBox} dataset can be used to address these challenges by facilitating speaker recognition research on singing voice data. Additionally, the dataset is annotated for language and gender labels, which can be used to investigate their effects on singer recognition performance. In the future, we plan to extend this dataset to include spoken voice audios for the singers in the current dataset. This will help us study the relationship between the spoken voice and the singing voice of a subject, in the context of speaker recognition.
\vspace{-0.2cm}
\section{Acknowledgements}
We thank Dr. J. P. Campbell from MIT Lincoln Laboratory for his valuable suggestions. We also thank Jefferson Bailey and Alexis Rossi from the Internet Archive for assisting with data collection for the \textit{JukeBox} dataset.

\bibliographystyle{IEEEtran}
\bibliography{main}

\atColsBreak{\vskip5pt} 
\end{document}